\documentclass[useAMS,usenatbib]{mn2e}

\usepackage{lscape}

\title[Mid-infrared excesses in G/K giants]{The incidence of mid-infrared excesses in G and K giants}
\author[M.H.Jones]{Mark H. Jones \thanks{e-mail: m.h.jones@open.ac.uk}\\
Department of Physics and Astronomy, The Open University, Walton Hall, Milton Keynes, MK7~6AA}

\begin{document}

\date{in original form 2008 February 6}

\pagerange{\pageref{firstpage}--\pageref{lastpage}} \pubyear{2008}

\maketitle

\label{firstpage}

\begin{abstract}
Using photometric data from the 2MASS and GLIMPSE catalogues, I investigate the incidence of
mid-infrared excesses ($\sim 10~\umu$m) of G and K stars of luminosity class III. In order to 
obtain a large sample size, stars are selected using a near-IR colour-magnitude diagram.
Sources which are candidates for showing mid-IR excess are carefully examined and modelled to 
determined whether they are likely to be G/K giants. It is found that mid-IR excesses are present at 
a level of $(1.8 \pm 0.4) \times 10^{-3}$. While the origin of these excesses remains uncertain,
it is plausible that they arise from debris discs around these stars. I note that the measured
incidence is consistent with a scenario in which dust lifetimes in debris discs are determined by 
Poynting-Robertson drag rather than by collisions. 
\end{abstract}


\section{Introduction}
It has long been recognised (e.g. Jura 1999) that the presence of infrared excesses in giant stars which have 
yet to evolve up the asymptotic giant branch may be due to the presence of long-lived debris discs around these stars.
Recent work on debris discs has concentrated on their properties during the main-sequence phase of 
stellar evolution but is becoming clear that some debris discs survive into the post-main sequence lives of stars,
as exemplified by the white dwarfs G29-38 and GD~362 \citep{b25,b01}, and possibly around the 
central star of the Helix Nebula \citep{b19}. The presence of detectable discs around evolved stars may
provide useful information about processes in debris discs since they have radiation environments which are distinct 
from systems in which the host star is on the main-sequence. The study of debris 
discs might be especially fruitful for giant stars which have yet to ascend the asymptotic giant branch, since
infrared emission is unlikely to be contaminated by emission from dust which has been 
lost from the star itself. 

To date, several studies have been made into the incidence of infrared excesses in giants. In the main, these 
studies have concentrated on the far-IR emission from giant stars. An important result from 
\citet{b17}, which is based on detailed analysis of data from the \emph{IRAS} Faint 
Source Catalog \citep{b16}, is that the incidence of 60 $\umu$m excesses in giants of 
spectral type G and K is $14 \pm 5$ per~cent. 

In the mid-IR (i.e. wavelengths of order 10 $\umu$m), the incidence of excesses in G/K giants 
is not well determined. There have been indications that there are some such stars which 
do exhibit mid-IR excesses. For example, a study by \citet*{b04},
identifies four stars with spectral type G or K and with luminosity class III which show an excess at wavelengths 
of $8~\umu$m. Another example of a G/K giant which shows mid-IR excesses is HD~233517, which was 
subject to mid-IR spectroscopic observations using \emph{Spitzer} by \citet{b13}, who concluded that 
this is likely to be a binary system with a peculiar history. 

An indication that some G/K giants may have mid-IR excesses is also evident from 
a study of mid-IR extinction by \citet{b11}. These authors noted, in passing, 
that in a selected sample of red-clump stars (which correspond to G/K giants) there 
appeared to be a small proportion of stars which exhibit 
excesses at $8~\umu$m in comparison to measured $K_{S}$ magnitudes (see their figure~4).
Indebetouw et al. attributed these mid-IR excesses to circumstellar dust but did not examine this population 
in detail. In particular, little attention was paid by these authors as to whether the stars with mid-IR excess
are actually G/K giants. In this work I investigate a similarly selected population of stars, but take considerable 
care in the selection of such a population and to pay particular attention to determining 
the nature of stars which show mid-IR excesses.

The aim of this study is to determine the incidence of excesses at 8$~\umu$m
in G and K giants, and to use available near- and mid-infrared photometry to investigate 
the nature of the sources showing such excesses. The most straightforward approach to such a study would be to 
cross-correlate a large spectral class catalogue, such as the Tycho-2 Spectral Catalog \citep{b23}
with a mid-IR catalogue such as the GLIMPSE Catalog \citep{b02}. This approach 
was investigated but was found to be limited by the 
fact that the number of spectroscopically identified G/K giants returned by such a cross-matching is only 
of order of a few hundred, and of these, many are relatively bright and hence are saturated in one or more photometric 
bands of the \emph{Spitzer} IRAC instrument. So, while this approach is satisfactory for investigating mid-IR 
excesses in main-sequence stars (and is the approach adopted by \citealt{b20}) it
returns too few matches to provide any useful information on the incidence of such excesses in G/K giants.

The approach adopted here is to use a near-IR 
colour-magnitude diagram to select a large sample (about $10^4$) of red-clump stars, and then to 
investigate in detail the stars from this sample which exhibit mid-IR excesses. In particular, I attempt to 
ascertain whether their spectral energy distributions are consistent with being G/K giants with excesses due to 
circumstellar dust and to rule out other plausible types of source.

This paper is structured as follows: Section 2 describes the data sets and the survey region adopted 
for this study and how a sample of red-clump giants was selected. Section 3
reports on the subset of these sources which are identified as having a mid-IR excess.
The implications of the observed incidence of genuine mid-IR excesses in G/K giants is discussed in 
Section 4. 

\section{Data sets and survey region}
The basis of this study is to use a near-infrared colour-magnitude (CM) diagram to select a 
population of objects which have a high probability of being G or K giants. A distinctive 
feature of any deep CM diagram of a sample which contains evolved stars of near-solar metallicity is the 
presence of a feature arising from the so-called red-clump, which corresponds to the core-helium burning phase of 
intermediate mass stars. Studies of the properties of red-clump giants, such as by \citet*{b26}
identify such stars as being of spectral type G or K and of luminosity class III (I shall use the 
terms red-clump stars and G/K giants interchangeably here). Stars which can be identified as belonging 
to this population have been used in a variety of contexts as standard candles; for instance, to study 
Galactic structure \citep{b15}. In this study, the identification of red-clump 
stars is used as a basis for studying the properties of the stars themselves. In particular, 
the near-IR CM diagram is used to select a population of stars for which near and mid-IR photometry is available. 

A common choice of photometric bands which can be used to construct the required CM diagram is
the near-IR $J$ and $K$ bands. In this study I use data from the 2MASS Point Source Catalog \citep{b18},
which comprises $J$, $H$ and $K_{S}$ photometric bands ($K_{S}$ overlaps with the standard $K$ band, 
but is slightly narrower and has a centre-of-band wavelength of 2.15$~\umu$m). Figure~1 shows 
the $(J-K_{S})$ versus ${K_{S}}$ colour magnitude diagram for the chosen survey region (a description of 
which is given below). In this diagram, the red-clump stars form the band of sources stretching 
from $(J-K_{S})\approx 1.0$, $K_{S}\approx 11$, down and to the right towards 
$(J-K_{S})\approx 1.8$, $K_{S}\approx 14$.

\begin{figure*}

\includegraphics{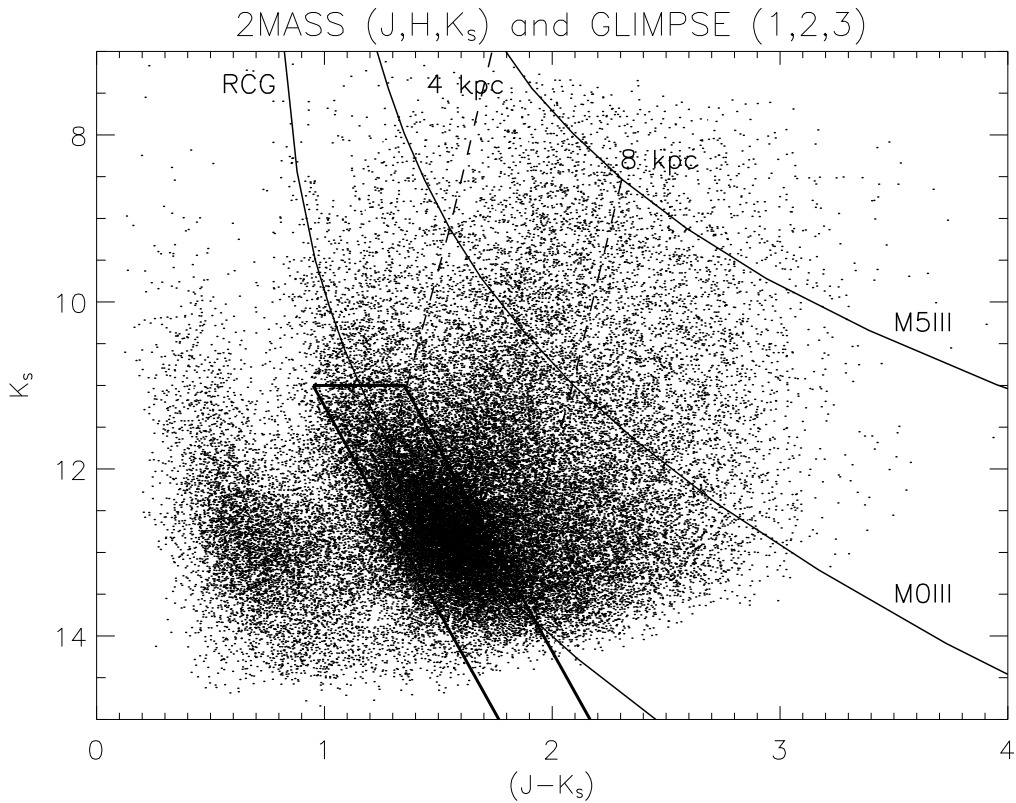}

\caption{The $K_{S}$ versus $(J-K_{S})$ colour-magnitude diagram of the survey area. Sources
included in this diagram are those 
which have photometric data in 2MASS at $J$, $H$, and $K_{S}$ and GLIMPSE at 3.6, 4.5 and 5.8$~\umu$m).
The solid curves indicate the loci expected for red-clump giants (RCG) and M giants under the 
assumption of uniform extinction per unit distance in the $K_{S}$ band of 0.09~mag~kpc$^{-1}$, with distances 
of 4 and 8~kpc indicated by dashed lines. The parallelogram (heavy lines) indicates the criteria used to 
select red-clump sources from this sample (see text for details).
}
\label{figure:CMD}
\end{figure*}

The identification of this feature with G/K giants is supported by its proximity to the expected locus 
of red-clump stars given a reasonable model for extinction along lines of sight in the survey region. 
If it is assumed that the extinction per unit distance in the $K_{S}$ band, $c_{K}$, 
is uniform, then for a given value of $c_{K}$, the tracks followed by different spectral classes of stars 
can be determined. Adopting the values given by \citet{b15} for canonical red-clump giants, 
that the absolute $K_{S}$ magnitude $M_{K}=1.65$ and the un-reddened colour $(J-K_{S})_{0}=0.75$, 
it is found that a good match between the red-clump feature and the expected location of 
such stars is obtained for $c_{K}\approx 0.09\mbox{~mag~kpc}^{-1}$ (indicated by the track labelled 
`RCG' in Figure~\ref{figure:CMD}).
This value of $c_{K}$ is in good agreement with the extinction data of \citet{b28} for the
chosen survey region when determined over scales of several kiloparsecs. 

 To the left and slightly below this feature
(which I refer to as the red-clump band), the sources which have $(J-K_{S})$ in the range
from 0 to about 1, are likely to be main sequence 
stars \citep{b27}, with later type stars showing relatively less reddening due to their proximity
to the Sun. 
To the right of the red-clump band the 
CM diagram is populated by giants ascending the giant branch and the asymptotic giant 
branch. The degree of discrimination between red-clump giants and later type giants is illustrated in
Figure~\ref{figure:CMD}, which also shows the expected loci of stars of type M0III and M5III 
(with $(J-K_{S})$ colours as given by \citealt{b27}, and absolute magnitudes derived from values given by 
\citealt{b29}) under the assumption of the same extinction model as 
adopted for the displayed locus of RCG stars. 
It can be concluded that the red-clump is a clearly defined feature in the colour-magnitude diagram. 

The survey area for this study was selected such that mid-IR photometry could be obtained from the 
GLIMPSE Catalog \citep{b02}\footnote{All GLIMPSE data products were obtained from http://www.astro.wisc.edu/sirtf/glimpsedata.html}. 
GLIMPSE is a survey of the Galactic plane ($10^{\circ}\leq|l|\leq65^{\circ}$ 
and $|b| \leq 1.0^{\circ}$) made using the four wavelength bands (3.6, 4.5, 5.8 and 8.0$~\umu$m) of the IRAC
instrument on \emph{Spitzer} \citep{b09}. A preliminary study for this work was based on the Version 1 release of the GLIMPSE
catalogues (released April 2005), while the results presented here are based on the Version~2 release (April 2007) unless 
otherwise indicated.

Any direction in the GLIMPSE survey region has a long 
line-of-sight through  the disc of the Galaxy. This is advantageous in terms of the numbers of red-clump 
stars that can be detected, but, of course, the survey regions are subject to relatively high levels of extinction.
Such extinction, corresponding to column densities of order $\mbox{a few} \times 10^{21} \mbox{~H atoms cm}^{-2}$, 
is not an issue as such, but it is important to appreciate that variation of extinction across a survey 
region will lead to a loss of definition (i.e. a broadening) of the red-clump band in the 
resulting CM diagram. 
In choosing any extended survey area, some variation of extinction is inevitable, but
this variation can be reduced by excluding regions with a high content of molecular gas.
In choosing the survey region, the aim was to maximise the area covered 
subject to the constraint of avoiding regions of high column density in molecular gas. 
This was done by avoiding known star formation regions, and by choosing a region of relatively 
low column density in molecular hydrogen from inspection of the CO emission map of 
\citet*{b06}.  
The region used for this study is defined by 
$45.0^{\circ} \leq \mathcal(l) \leq 50.25^{\circ}$ and 
$0.70^{\circ} \leq \mathcal(b) \leq 1.0^{\circ}$. 
For the area covered by the GLIMPSE survey, the selected region has relatively low column 
densities of CO.

The number of 2MASS sources with measured values of $J$ and $K_{S}$ in 
the selected region is 93888. The extent of the survey area is such that about $10^4$ clump giants 
would be included in the final sample, which is sufficient to measure the incidence of mid-IR excesses at a 
level of order 0.1 per~cent.

\subsection{Selection of red-clump sources}
As noted above, the selection of red-clump sources is based on the ($J-K_{S}$)-$K_{S}$ CM diagram. 
Rather than using all sources with 2MASS $J$ and $K_{S}$ data, preliminary studies 
revealed that an improvement in the degree of contrast between the red-clump and the main 
sequence features can be achieved by requiring that included sources are detected
the all 2MASS bands and the 3.6, 4.5 and 5.8 $\umu$m bands of the GLIMPSE Catalog.  
The number of sources in the survey region which satisfy this criterion is 53520 and it is
these sources which are shown in the CM diagram in Figure~\ref{figure:CMD}.

Stars which have a high probability of being G/K giants 
were selected by following the leftmost side of the the red-clump band which is evident in 
Figure~\ref{figure:CMD}. While such a selection is somewhat subjective, 
using the left-hand side of the red-clump distribution in this way
should minimise contamination of the sample from stars ascending the 
giant branch/asymptotic giant branch. 
The selection criterion 
used was that $K_{S} \geq 11.0$ and that $(J-K_{S})$ 
should lie within $ \pm 0.2$
magnitudes of the straight line defined by 
\begin{equation}
(J-K_{S}) = 1.15 + 0.204 (K_{S} - 11.0)
\label{eqn:redclump}
\end{equation}
This selection region is overlaid on Figure~\ref{figure:CMD}. Note that the cutoff at 
$K_{S}=11.0$ was determined by the fact that at lower magnitudes the locus of the red-clump band becomes 
difficult to trace due to the small number of sources in this part of the CM diagram. 
It can be seen from Figure~\ref{figure:CMD} that this 
selection region is only an approximation to the expected track followed by red-clump stars 
under a model of uniform extinction per unit distance, but is a reasonable method of selecting such stars 
in the interval $11.0 < K_{S} < 13.5$ (in fact, the faint cut-off adopted in this work is brighter
than $K_{S}=13.5$, see below). 

The CM diagram and selection region shown in Figure~\ref{figure:CMD} allows an estimate to be made 
of the maximum amount of contamination of the red-clump region by main-sequence stars. The profile of
the distribution at constant $K_{S}$ was examined at $K_{S}\approx 13$. The profile reveals 
a bimodal distribution with peaks due to main-sequence and evolved stars. 
At the minimum between the peaks of this
distribution, the density of main-sequence sources is about 10~per cent of 
the source density in the red-clump band. Since this minimum lies somewhat to the left
of the red-clump band, and assuming that the density of main-sequence stars decreases 
as ($J-K_{S}$) increases, the contamination of main-sequence stars in the selected
region will be less than 10~per cent: a reasonable estimate on an upper limit to 
the contamination would be at the level of a few percent.

In addition to adopting selection criteria which isolate red-clump stars, it is also important to  
choose a faint cut-off in the value of $K_{S}$. This serves to reduce the bias 
inherent to detecting excesses at $8 ~\umu$m when many stars are close to the detection threshold at this 
wavelength. Ideally, it would be best to select a cut-off in $K_{S}$ which results in all 
sources being detected at $8 ~\umu$m. However, such a constraint leads to the rejection of a large 
proportion of the data. I investigated (using Version~1 of GLIMPSE) how the detectability of red-clump 
sources at $8 ~\umu$m varies with $K_{S}$ and found that at $K_{S}=12.9$, 85 per~cent of sources 
were detected. This cut-off in $K_{S}$ was adopted as being a reasonable compromise between expected 
detection and source numbers. With these criteria applied ($11 \leq K_{S} \leq 12.9$ and Equation~\ref{eqn:redclump})
the number of red-clump sources selected is 9865. This sample forms the basis of the study of mid-IR excesses.

\subsection{The selection of sources with $8 ~\umu$m excess}
\begin{figure*}

\includegraphics{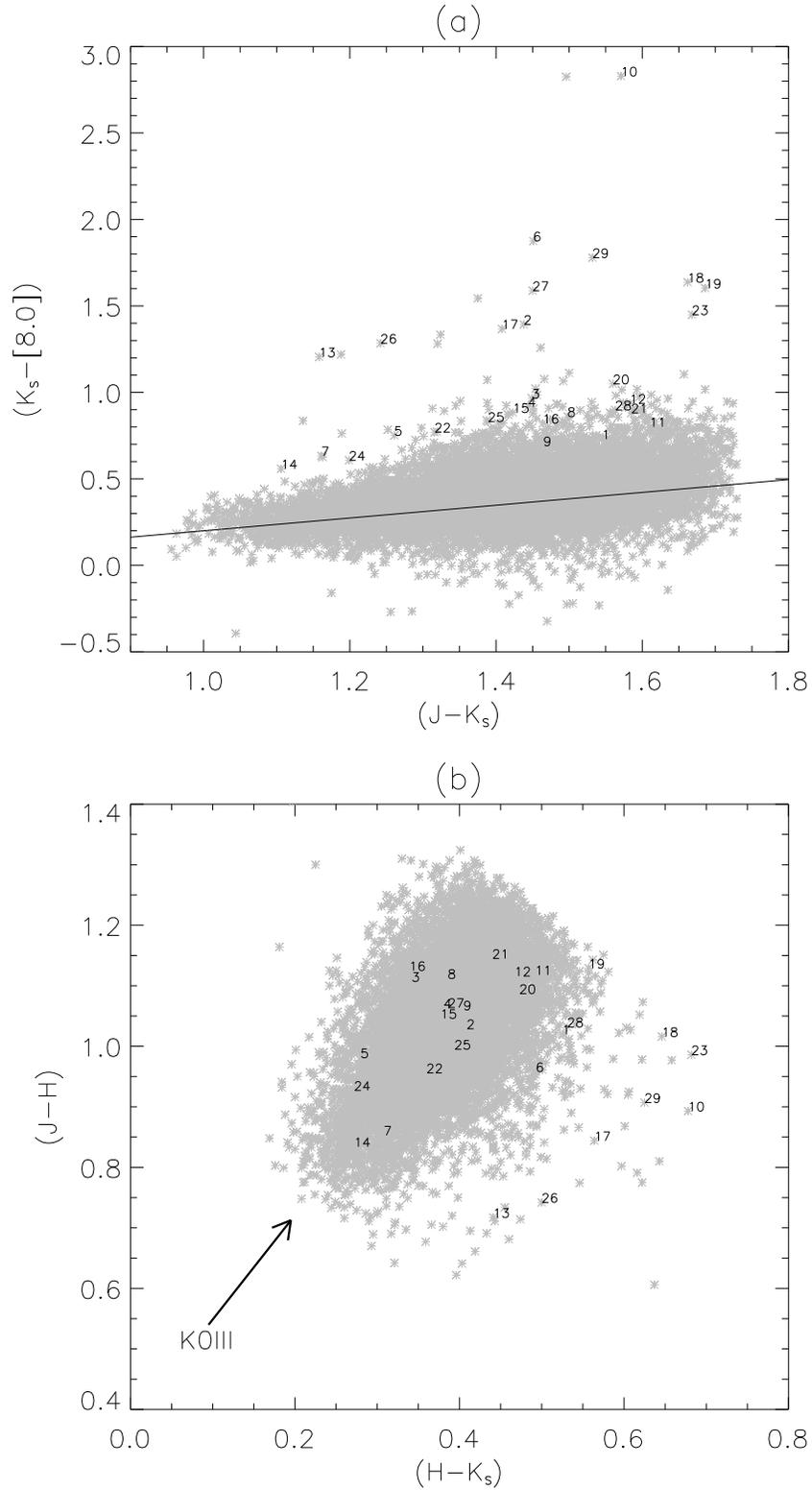}

\caption{Colour-colour diagrams(a) $(K_{S}-[8.0])$-$(J-K_{S})$, and (b) $(J-H)$-$(H-K_{S})$, 
for all of the selected red-clump stars (grey asterisks).
Candidate sources with excesses at 8~$\mu$m are indicated by the numerical identifier as given in 
Table~\ref{table:candidates}.  The straight line in (a) shows the extinction relation 
as derived by \citet{b11} using a similar sample of stars. 
The reddening vector shown in (b) starts at the expected position of an un-reddened star of type 
K0III.}

\label{figure:JH_HK}
\end{figure*}

An indication of the incidence of sources with excesses at $8~\umu$m in the sample is revealed by the 
$(K_{S}-[8.0])$-$(J-K_{S})$ colour-colour diagram of the selected sample of red-clump sources 
(Figure~\ref{figure:JH_HK}a) . 
It can be seen that there is a small number (about 20) of sources with excesses in $(K_{S}-[8.0])$ 
which set them clearly apart from the main distribution of sources. Sources with less pronounced, but still significant 
excesses, can only be identified from more detailed analysis as described below. 

It can also be seen from Figure~\ref{figure:JH_HK}a that the 
effects of  interstellar reddening are apparent in the distribution of sources, with a trend for $(K_{S}-[8.0])$ to 
increase with increasing $(J-K_{S})$. It is therefore useful to define a colour $(K_{S}-[8.0])_{0}$ which includes the effects of 
reddening.
The results of Indebetouw et al. (2005), and in particular, their best-fitting straight line
to ($[8.0]-K_{S}$)-($J-K_{S}$) data (as shown in their figure~4a and reproduced 
here on Figure~\ref{figure:JH_HK}a), allows this to be calculated as
\begin{equation}
(K_{S}-[8.0])_{0} = (K_{S}-[8.0])-0.37 \times (J-K_{S})+0.17
\end{equation}
For the sample of red-clump sources described above, the mean value, 
$\langle (K_{S}-[8.0])_{0} \rangle$ is $0.048 \pm 0.002$. 
The colour excess of any source is given by a colour excess defined by 
$ E_{K8}=(K_{S}-[8.0])_{0}- \langle (K_{S}-[8.0])_{0} \rangle $

Sources which are candidates for stars having excesses at $8 ~\umu$m were selected using the 
criterion
\begin{equation}
\mbox{SNR}_{EK8}=\frac{E_{K8}} {\sigma_{E_{K8}}} > 3.7 
\end{equation}
The significance level here is determined by the sample size: it is chosen such 
that given a Gaussian distribution in $E_{K8}$, it would be expected that there would be only one 
source in a sample size of $1 \times 10^{4}$
which would exceed this value. On applying the above criterion, 52 sources were identified as 
candidate mid-IR excess sources.

\section{Analysis of candidate $\hbox{8~$\umu$\lowercase{m}}$ excess sources}
In order to confirm or reject the candidate excess sources, the following approach was taken. Firstly,
where doubts exist over the reliability of a source, it was culled from the sample. The 
remaining sources were then analysed in two ways: on the basis of their near-infrared colours
and by modelling of the photometric data. 

\subsection{Removal of low reliability sources}
The first stage in checking source reliability was to inspect the GLIMPSE Catalog source quality flag (SQF) 
for each source. Of the 52 candidate sources, 14 had SQF flags set which indicated 
confusion in in-band or cross-band merging. Two further sources were flagged as being potentially 
affected by banding correction and a single source was flagged as being possibly contaminated by stray light. 

In addition to the SQF data provided in the GLIMPSE Catalog, the $8 ~\umu$m GLIMPSE Image Atlas 
(Version 2) was visually inspected to check for artefacts which might have given rise to anomalous 
$8 ~\umu$m fluxes. The highest resolution (0.6 arcsec pixel size) $8 ~\umu$m GLIMPSE atlas images were used to 
check for anomalies. Fields of extent 1.2 arcmin around source positions were inspected and six sources (which
would have been accepted on the basis of their SQF value of zero) were 
rejected on the basis of their appearance. Four of these sources appear to be affected by stray light 
contamination at $8 ~\umu$m, while two are very close to relatively bright multiple sources. All six of these
sources are also deemed to be of questionable reliability and, together with the 17 sources identified 
by their SQF values, were rejected from the candidate list. The 2MASS and GLIMPSE photometry for the 
remaining 29 candidate sources is given in Table~\ref{table:candidates}. Note that hereafter, 
individual sources are referred to by the identifier (ID.) given in this table.

\begin{table*}
 \vbox to220mm{\vfil Landscape table to go here.
 \caption{}
\vfil}
\end{table*}

\subsection{Near-IR colours}
Some evidence as to the nature of the stars exhibiting 8~$\umu$m excesses can be determined by inspection of
the $(J-H)$-$(H-K_{S})$ colour-colour diagram (Figure~\ref{figure:JH_HK}b). The locations of the candidate 
8~$\umu$m excess sources are indicated by the numerical identifier given in Table~\ref{table:candidates}. 
The vast majority of the sources in this diagram are distributed in a band which corresponds well 
to the expected location of G/K giants under different amounts of extinction (for reference, a reddening vector 
is shown which indicates the expected direction of the distribution of stars of type K0III). 
While the majority of candidate 8~$\umu$m excess sources lie within the region populated by reddened G/K giants, 
it is notable that seven of the candidate excess sources have $(J-H)$ values which are substantially lower 
than would be expected from their $(H-K_{S})$ colour (these are sources with identifiers 10, 13, 17, 18, 23, 26 
and 29). If these sources are G/K giants, they would appear to have anomalous near-IR colours. The remaining 
22 candidate sources appear to be consistent with being G/K giants. Further investigation (below) uses the 
available 7-band photometry available to model these sources.

\subsection{Modelling of 7-band photometry}
The 29 sources with excesses in their $(K_{S}-[8.0])_{0}$ colours were further analysed to 
attempt to determine if these sources were likely to be G/K giants with excesses at 
$8 ~\umu$m, or if they have some other origin. As noted above, the sample of red-clump stars is expected 
to be contaminated by other types of source at the level of maybe as much as a few percent. However, it 
could plausibly be the case that it is these contaminating stars which give rise to the sources 
detected with 8\mbox{$~\umu$m} excesses. Hence it is important then to consider whether the detected excess 
sources could be other types of star. The approach taken was to use the data from the seven photometric bands 
($J$, $H$, $K_{S}$, $3.6 \umu$m, $4.5 ~\umu$m, $5.8 ~\umu$m, and $8.0 ~\umu$m)
to compare to the spectra from a selection of plausible models. 

Clearly, one model which needs to be tested is that of a red-clump giant with a mid-IR excess. 
As a reference point, it is useful to define a `canonical' red-clump giant using the results of 
the study of \citet{b26}. 
The subset of metal rich ($[\mbox{Fe}/\mbox{H}]>-0.3$) clump giants in their study have 
a mean effective temperature of $T_{\mbox{eff}}=4860\mbox{~K}$ and surface gravity of $\log g = 2.81$. 
The canonical red-clump giant can be approximated by the closest Kurucz model (see below), 
which has $T_{\mbox{eff}}=4750 \mbox{~K}$, and $\log g =3.0$.

In model fitting however, I adopted a range of giant star models with a range of 
surface temperatures from 3500~K to 6500~K. This is a wider temperature range than would be expected
for red-clump giants, but provides a useful check that individual stars are not better modelled as 
cooler or hotter stars.  This, and all other
photospheric emission (in other models described below) was modelled using a Kurucz 
spectral model\footnote{Obtained from http://wwwuser.oat.ts.astro.it/castelli/grids.html}
\citep{b03} with solar metallicity (also, the particular set of 
models used have a turbulent velocity of
$2 \mbox{ km s}^{-1}$). The temperature resolution of the published grid of Kurucz models 
is 250~K.  

The mid-IR excess is modelled as a component which only contributes 
significantly to the $8 ~\umu$m band. This was implemented as a Gaussian component at 
a wavelength of $7.9 ~\umu$m with $\sigma =0.6 ~\umu$m, with only the normalisation of this 
parameter being allowed to vary. While this component is a reasonable representation
of a PAH feature observed in the spectrum of HD~233517 
\citep{b13}
it should be
stressed that an excess in one photometric band could be modelled in a variety of ways, 
and that I do not claim any particular significance to modelling the excess in this way.

A quite different set of stars which are known to show mid-IR excesses are hot stars 
with infrared excesses arising from bremsstrahlung emission from plasma surrounding the 
star \citep{b10}. Such sources are detected in considerable numbers by \citet{b04} and
\citet{b20}
and are believed to occur for stars of spectral type B8 or earlier. Consequently I also examine a
hot star model comprising a main-sequence B8 star and bremsstrahlung emission from a plasma at a 
temperature of $1 \times 10^{4} \mbox{K}$. Note that as far as fitting to the 
combined 2MASS and GLIMPSE photometry is concerned, the model is rather insensitive to the exact 
stellar type: the choice of B8V is a conservative one which represents the coolest and lowest 
luminosity star that might plausibly also exhibit an infrared excess due to bremsstrahlung emission.

A component of all of the fitted models was the inclusion of extinction due to 
interstellar dust. The extinction values used were taken from the publicly available 
synthetic extinction curves\footnote{http://www.astro.princeton.edu/\~{ }draine/dust/dustmix.html} 
which are based on the grain model of \citet{b14} and the grain size distribution of \citet{b22}.
The particular extinction curve
used was the one for which $R_{V}=3.1$ and which these authors consider to provide 
a good match to the extinction properties of diffuse HI clouds in the Milky Way.

For all of the models, three parameters were fitted: the normalisation $X$ of 
the stellar spectral component, the normalisation of either the bremsstrahlung component or of the 
Gaussian line at $7.9 ~\umu$m, and the line-of-sight column density $N_{H}$. In addition to these
parameters, note that for the red-clump giant model, the fitting was conducted across a range temperatures 
(as described above) in order to also determine the photospheric temperature.
A spectrophotometric distance 
$d$ to the star can be derived from the normalisation parameter $X$ by 
$d=r_{\ast}X^{-1/2}$ where $r_{\ast}$ is the stellar radius. Stellar radii were estimated 
by assuming blackbody emission and relationships between stellar luminosity and temperature as 
given by \citet{b05}. 

It is important to appreciate that the spectral resolution 
provided by the seven photometric bands used here is, in general, insufficient to 
provide any clear discrimination in luminosity class: data which are well modelled as a 
G/K giant with a Gaussian excess at $8 ~\umu$m, will also be acceptably fitted by models in which the 
star is a G/K dwarf or supergiant. Models in which the star is modelled as a dwarf and as a supergiant 
were also fitted to obtain spectrophotometric distances and values of column density for the excess sources. 
In order to investigate whether these models could be plausible, a set of
neighbouring (non-excess) red-clump giants was used to estimate the column density in the direction of 
each candidate excess source. The set of neighbouring sources for each candidate excess source was defined 
as those sources which were selected according to the red-clump criterion (Equation~\ref{eqn:redclump}) 
and which lie within 5~arcmin of the position of the excess source and which have $K_{S}$ within 0.2 mag of this 
source. In fitting to these neighbouring  red-clump giants, the canonical red-clump stellar 
model was assumed (i.e. $T_{\mbox{eff}}=4750 \mbox{~K}$, and $\log g =3.0$). The fitting process returned 
a column density and flux normalisation ($X$) for each source, allowing a mean column density and mean 
spectrophotometric distance for the neighbouring red-clump stars to be determined. 
These values for a so-called `neighbouring field' are given in Table~\ref{table:fitting} and are discussed 
further in Section~3.4.

\setcounter{table}{1}
\begin{table*}
\small
\begin{tabular}{rlcrrccrrcr}

&                          & \multicolumn{7}{c}{Best fit model} & \multicolumn{2}{c}{Neighbouring field} \\ 
\multicolumn{1}{c}{ID.}
 & \multicolumn{1}{c}{Designation}              
                   & \multicolumn{1}{c}{Type}
                        & \multicolumn{1}{c}{$T/\mbox{K}$}
                                & \multicolumn{1}{c}{$\chi^2_{\nu}$}
                                        & \multicolumn{1}{c}{$N_{H}$}
                                               & \multicolumn{1}{c}{$d / \mbox{kpc}$}
                                                        & \multicolumn{1}{c}{$E_{8}$}
                                                             & \multicolumn{1}{c}{$\mbox{SNR}_{8}$} 
                                                                  & \multicolumn{1}{c}{$N_{H}$ }
                                                                                  & \multicolumn{1}{c}{$d / \mbox{kpc}$} \\ \hline
 1 & G045.1000+00.8808 & RCG & 5750  & 0.69 & $12.7\pm0.5$ & $2.1\pm0.4$ &  0.13 & 2.3  & $8.3\pm1.0$ & 4.4\\
 2 & G045.1844+00.7612 & RCG & 5000  & 0.33 & $ 9.8\pm0.9$ & $4.4\pm1.0$ &  0.89 & 9.2  & $9.1\pm1.1$ & 5.4\\
 3 & G045.2769+00.7123 & RCG & 4500  & 0.40 & $ 7.9\pm0.9$ & $6.7\pm1.7$ &  0.52 & 3.9  & $9.1\pm1.1$ & 5.4\\
 4 & G045.2852+00.9988 & RCG & 4500  & 0.41 & $ 7.7\pm1.0$ & $7.0\pm2.1$ &  0.52 & 5.0  & $9.2\pm0.8$ & 5.5\\
 5 & G045.4108+00.8768 & RCG & 4500  & 0.38 & $ 5.8\pm0.9$ & $5.9\pm1.5$ &  0.42 & 5.7  & $7.4\pm2.4$ & 4.6\\
 7 & G046.2397+00.9526 & RCG & 5000  & 0.13 & $ 6.6\pm0.6$ & $2.7\pm0.4$ &  0.28 & 5.9  & $6.8\pm1.3$ & 3.3\\
 8 & G046.3252+00.8710 & RCG & 4750  & 0.58 & $ 9.6\pm0.7$ & $5.4\pm1.1$ &  0.33 & 4.2  & $9.2\pm1.1$ & 5.5\\
 9 & G046.6118+00.9688 & RCG & 4500  & 0.81 & $ 8.0\pm0.9$ & $5.2\pm1.3$ &  0.27 & 4.0  & $8.3\pm0.9$ & 4.1\\
10 & G046.6781+00.7324 & HSB & 11500 & 0.66 &  15.2        & $>1.8$      &  2.45 & 82   &             &    \\
11 & G046.6833+00.8797 & RCG & 4500  & 0.90 & $ 9.3\pm0.9$ & $6.1\pm1.5$ &  0.35 & 4.6  & $9.6\pm0.9$ & 4.9\\
12 & G046.7233+00.7512 & RCG & 4250  & 1.05 & $ 7.9\pm0.8$ & $8.6\pm2.2$ &  0.54 & 4.5  & $9.8\pm0.7$ & 5.2\\
13 & G046.9192+00.7417 & HSB & 11500 & 0.65 &  13.1        & $>1.1$      &  0.65 & 19   &             &    \\
14 & G047.5359+00.8104 & RCG & 5750  & 1.96 & $ 8.3\pm0.6$ & $1.8\pm0.3$ &  0.12 & 2.7  & $7.6\pm1.0$ & 3.3\\
15 & G047.8477+00.9287 & RCG & 4000  & 1.36 & $ 4.9\pm0.6$ & $8.6\pm3.2$ &  0.64 & 5.4  & $7.6\pm1.0$ & 4.0\\
16 & G048.1801+00.9573 & RCG & 4250  & 0.26 & $ 6.9\pm1.1$ & $8.2\pm2.2$ &  0.43 & 4.3  & $8.4\pm0.9$ & 5.1\\
17 & G048.3706+00.7076 & HSB & 11500 & 1.24 &  15.3        & $>1.7$      &  0.75 & 11   &             &    \\
19 & G048.3924+00.8103 & RCG & 5000  & 0.78 & $12.1\pm1.1$ & $4.4\pm1.5$ &  1.04 & 12   & $9.5\pm1.3$ & 5.5\\
20 & G048.4911+00.8260 & RCG & 4500  & 0.96 & $ 9.2\pm1.1$ & $6.2\pm1.9$ &  0.57 & 4.0  & $9.0\pm1.1$ & 5.1\\
21 & G048.5964+00.7855 & RCG & 4500  & 0.15 & $ 9.2\pm0.8$ & $6.5\pm1.6$ &  0.40 & 3.9  & $9.7\pm1.1$ & 5.2\\
22 & G048.5970+00.9849 & RCG & 4750  & 0.74 & $ 7.2\pm0.8$ & $5.5\pm1.2$ &  0.40 & 4.0  & $8.4\pm1.1$ & 5.2\\
23 & G048.6036+00.9284 & HSB & 11500 & 0.51 &  17.7        & $>1.6$      &  0.75 & 12   &             &    \\
24 & G049.0793+00.7610 & RCG & 4500  & 0.57 & $ 5.0\pm1.0$ & $5.6\pm1.6$ &  0.33 & 4.6  & $7.4\pm1.2$ & 4.3\\
25 & G049.1099+00.8232 & RCG & 4500  & 0.59 & $ 7.0\pm0.9$ & $6.5\pm1.6$ &  0.48 & 5.2  & $8.7\pm1.1$ & 5.1\\
26 & G049.4210+00.8476 & HSB & 11500 & 0.60 &  13.5        & $>1.6$      &  0.76 & 11   &             &    \\
27 & G049.5812+00.9666 & RCG & 4250  & 0.30 & $ 6.8\pm1.0$ & $8.6\pm2.9$ &  1.23 & 7.2  & $8.0\pm0.9$ & 5.4\\
28 & G049.9290+00.8456 & RCG & 6250  & 0.32 & $14.0\pm0.6$ & $1.8\pm0.3$ &  0.23 & 3.0  & $9.2\pm1.0$ & 5.6\\
29 & G050.1256+00.9808 & HSB & 11500 & 0.57 &  16.3        & $>1.6$      &  1.15 & 26   &             & 
\end{tabular}
\normalsize
\caption{The results of model fitting. Column 1 shows the source identifier (ID.) as used in 
Table~\ref{table:candidates}. 
Column densities ($N_{H}$) are expressed in units of $(10^{21} \mbox{H atoms cm}^{-2})$.
The quantities $E_{8}$ and $\mbox{SNR}_{8}$ are defined in the text.} 
 
\label{table:fitting}
\end{table*}

For a particular model, the expected flux density in each photometric band was obtained by integrating uniformly over a
passband between the wavelengths at which the filter response is half its maximum value. 
Such an integration is an approximation to modelling the exact photometric responses of the filters, 
but is expected to be adequate for this work.

The fitting process was a minimisation of chi-squared type approach, using the MPCURVEFIT routine in IDL. 
Given that fitting was carried out at fixed photospheric temperatures, the uncertainties in the fitted
parameters which were returned by this process were unrealistically small. 
Consequently, uncertainties in fitted parameters (and quantities derived from these parameters)
were estimated by inspection of the parameters returned from a range of temperatures corresponding to 
variation in $\chi^{2}$ of about 1.

\subsection{Results of modelling}
The modelling process has two distinct aims. The first is to determine whether a given excess source can be 
acceptably fitted by a model which comprises a red-clump giant and Gaussian excess at $8 ~\umu$m  
(a model referred to here as RCG) or by a model comprising a hot star 
and bremsstrahlung emission (which is referred to as HSB). The second aim is to determine which of the 
those stars which are well modelled by the RCG model do actually exhibit a significant excess at $8 ~\umu$m.

\begin{figure*}

\includegraphics{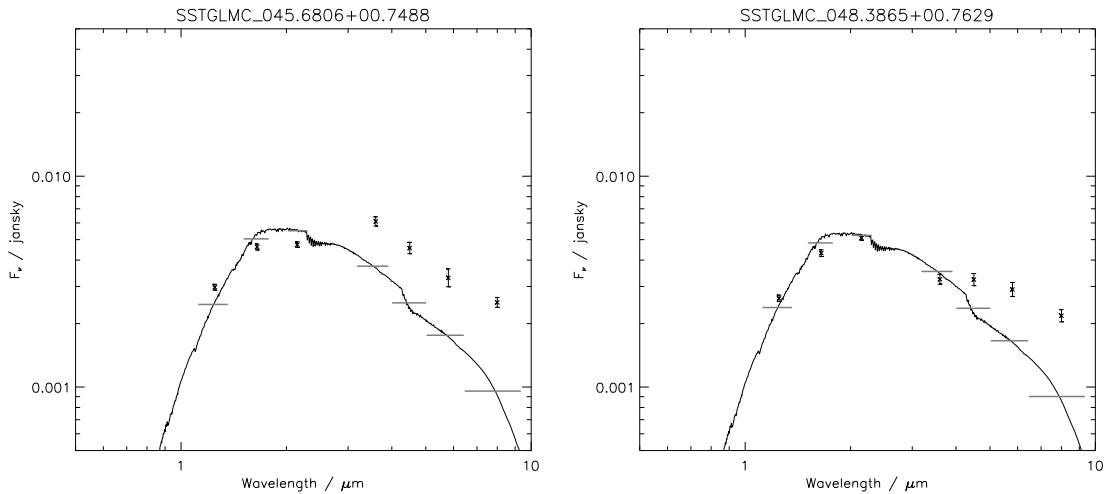}

\caption{The spectral energy distributions of the two candidate excess sources (6 and 18, 
shown on left and right respectively)
which are not well fitted by either the RCG or HSB models (see text). 
Photometric data points are 
indicated by '$\times$' with error bars. For comparison the spectrum of a canonical red-clump giant 
is also shown (continuous line) together with the model photometric points (grey bars - the width of 
which  represent the FWHM of the photometric bands).}
\label{figure:SED_BAD}
\end{figure*}

Of the 29 candidate sources, 27 could be fitted by either the RCG or HSB models within the 95~per~cent 
confidence interval and are taken to be acceptably modelled. 
The two sources (6 and 18) which were not acceptably fitted 
had minimum values of $\chi^{2}_{\nu} > 5$ and hence the models
can be rejected at 
the 99.9 percent confidence level. The spectral energy distribution of source 6 
(shown in Figure~\ref{figure:SED_BAD}, left-hand panel) suggests that it has a continuum level which differs
between the 2MASS and GLIMPSE observations. Since
2MASS and GLIMPSE data were collected at different times, a likely explanation is that this is a
variable star and it is discounted from any further analysis. 
The nature of source~18 is less clear: this could be an RCG source with infrared excesses at as short a wavelength as
4.5~$\umu$m, although its spectral energy distribution (Figure~\ref{figure:SED_BAD}, right-hand panel) 
also appears to be reasonably well-described by the 
HSB model with the exception of 
the flux measurement at 3.6~$\umu$m. While it is possible that variants to the RCG or HSB model could describe the data for
this source, such modelling is not justified using these data. It is also noted that this source is one of the seven sources
in Figure~\ref{figure:JH_HK}b which lie away from the expected distribution of G/K giants (see Section~3.2). 
The nature of source~18 is acknowledged to be uncertain and this 
source is excluded from further analysis. 

For all of the sources which can be acceptably fitted by either the RCG or the HSB models, an 
excess at $8 ~\umu$m (called $E_{8}$) was determined by calculating the difference between the observed 
value of [8.0] and the model photospheric value ($[8.0]_{\mbox{\scriptsize phot}}$). The signal-to-noise ratio of the 
excess was calculated as
\begin{equation}
\mbox{SNR}_{8}=\frac{[8.0]-[8.0]_{\mbox{\scriptsize phot}}}{\sigma_{[8.0]}}
\end{equation}
This quantity is tabulated, along with the results of the model fitting in Table~\ref{table:fitting}.

\begin{figure*}

\includegraphics{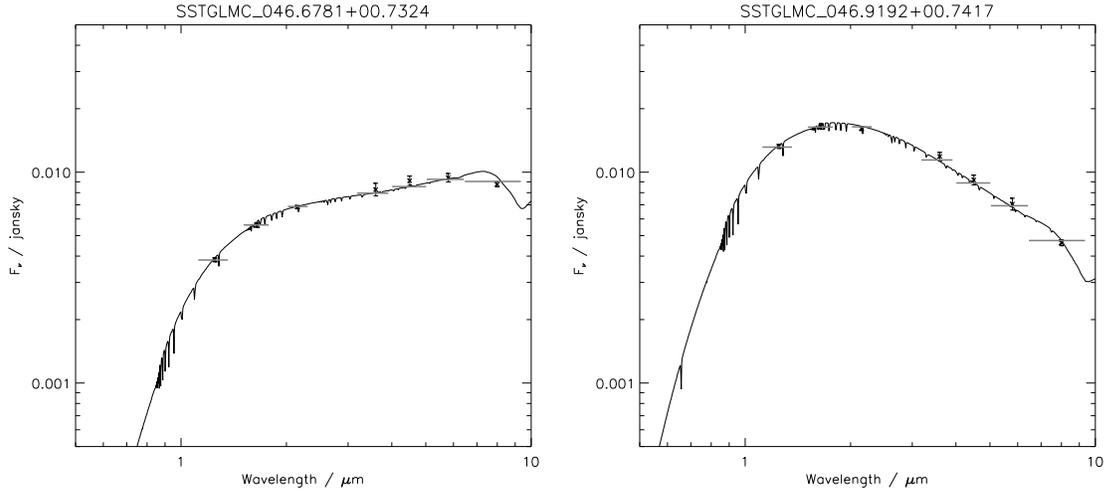}

\caption{The spectral energy distributions of two excess sources (10 and 13, shown 
on left and right respectively) which are  
well fitted by HSB models (see text). These two examples illustrate the wide range of 
continuum slope which can be generated by this model. 
The spectral energy distributions are displayed in the same manner as in Figure~\ref{figure:SED_BAD}.}
\label{figure:SED_HSB}
\end{figure*}

Of the 27 sources which are well modelled by either the HSB or RCG models, it should be noted that the 
distinction between these two classes of models is very strong: no source can be acceptably modelled 
by both types of model. Six of the 27 sources are better fitted by the HSB model than by the RCG model, 
and all of these sources were identified from Figure~\ref{figure:JH_HK} as being unlikely to be G/K giants. 
The
sources are characterised by high column densities ($>10^{22} \mbox{H atoms cm}^{-2}$) and very strong 
mid-IR excesses ($\mbox{SNR}_{8}>10$). Note that because the model assumes the star is the lowest luminosity star 
which is likely to exhibit bremsstrahlung emission (i.e. type B8V) the distances quoted in Table~\ref{table:fitting} 
are lower limits. Two examples of the spectral energy distribution of sources that are best fitted by the 
HSB model are shown in Figure~\ref{figure:SED_HSB}. It is noticeable that this model can reproduce a wide range 
of spectral slopes at $\lambda > 2 ~\umu$m. In particular, the source 10 seems to have spectral energy 
distribution which rises between the 2MASS and the GLIMPSE bands, and it is clear that this behaviour can be successfully 
described by the HSB model.

The remaining 21 sources are well-fitted by the RCG model, and it is notable that in 
Figure~\ref{figure:JH_HK}, all of these sources are at locations consistent with them being G/K giants. 
The quantity $\mbox{SNR}_{8}$ is used
to determine whether the excess at $8 ~\umu$m is confirmed. As in the case for selecting candidate excess sources 
using $E_{K8}$, the cut-off is at a level where only one excess source is expected to occur by chance from the 
total sample of red-clump giants, i.e. $\mbox{SNR}_{8}>3.7$.
Of the 21 candidate excess sources which can be fitted by the RCG model, 18 show excesses which satisfy 
this criterion, and of these, seven sources have $\mbox{SNR}_{8} \geq 5$. 

Three sources (1, 14, and 28) do not have confirmed excesses.
It is noticeable that these are all of high photospheric temperature ($T_{\mbox{eff}}=5750 \mbox{~K}, 6250 \mbox{~K}$).
Such temperatures would be unusually high for a red-clump star, and it is possible that these stars are
low metallicity stars on the horizontal branch. However, given that their mid-IR excesses are unconfirmed, 
they are rejected from the sample. 

\begin{figure*}
\includegraphics{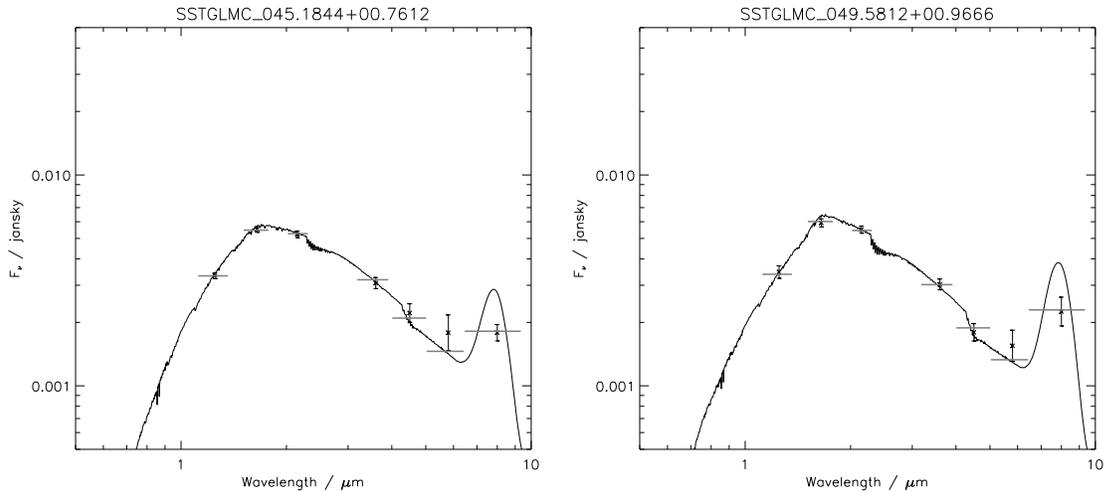}

\caption{The spectral energy distributions of two excess sources (2 and 27, shown on left and right respectively) which are  
well fitted by RCG models (see text). These are the two sources 
of this type with the strongest 8~$\umu$m excesses.
The spectral energy distributions are displayed in the same manner as in Figure~\ref{figure:SED_BAD}.}
\label{figure:SED_RCG}
\end{figure*}

The remaining 18 sources are strong candidates for being G/K giants with mid-IR excess. Examples of 
the spectral energy 
distributions of two of these sources are shown in Figure~\ref{figure:SED_RCG}.  
As might be expected from the goodness-of-fit
for these sources, the RCG model provides an excellent match to these data, with no indication of 
systematic deviations which might cause us to question the validity of the model. However, one possibility 
that should be addressed is that these stars might be of a different luminosity class to that assumed in the 
model, i.e. that the stars may be on the main-sequence or supergiants.

This was investigated by fitting using Kurucz models corresponding to G/K dwarfs and supergiants
(using Version 1 of the GLIMPSE data). In the case of models which are well-fitted as red-clump giants, 
the best-fitting dwarf or supergiant model typically has a very similar temperature (usually differing by no more than 250~K) 
and similar column density to the best-fitting giant star model. The distance estimates, as would be expected, 
depend very strongly on the assumed luminosity class. As can be seen from Table~\ref{table:fitting} the distances implied by 
giant star models are broadly consistent with those obtained from the neighbouring red-clump stars (labelled as the `neighbouring 
field' in Table~\ref{table:fitting}). Dwarf star models 
(luminosity class V) imply distances which are typically an order of magnitude lower than the mean 
distance to the neighbouring red-clump giants, but at a similar value of total column density. 
It is concluded that it is highly unlikely that any of the candidate excess stars 
are on the main-sequence. The distance estimates which are arrived at assuming that the candidate 
excess star is a supergiant of luminosity class I are typically of order 100~kpc. The assumption that 
these stars are supergiants can be rejected since such distances would place these stars well outside the 
stellar disc of the Milky Way. It seems reasonable to conclude that the candidate excess stars are 
all giants (i.e. luminosity class III).

\section{Discussion}
The analysis presented here confirms that there is a population of red-clump stars which appear to be 
G/K giants with significant mid-IR excesses. Out of an initial sample of 9865 red-clump stars, the analysis
presented here identifies 18 sources which fall into this category. The incidence of mid-IR excesses in 
G/K giants is thus calculated to be $(1.8 \pm 0.4)\times 10^{-3}$. This is substantially lower than 
the incidence of far-IR excesses in G/K giants of $14 \pm 5$ percent reported by \citet{b17}. 
This low-incidence of mid-IR excesses as compared to far-IR excesses mirrors, in a qualitative sense at least, 
the behaviour seen for main-sequence stars, where mid-IR excesses are far less common than those detected at 
far-IR wavelengths, which is interpreted as reflecting the incidence of systems with hot ($\sim \mbox{few} \times 100$~K)
and cold ($ < 100$~K) dust respectively \citep{b20}. A discussion of implications of our observed 
incidence of mid-IR excesses in G/K giants is given below. 

In investigating candidate excess sources, this work has also identified a substantial number of sources
whose SEDs are well-described as highly absorbed early-type 
stars in which the excess emission arises from bremsstrahlung emission (the HSB model). 
With the admittedly limited information provided by the 7-band photometry used here, some caution 
should be expressed about the physical interpretation of this model: it may plausibly be the case that 
this type of SED is produced by a G/K giant (or other luminous star) which has infrared excesses 
across most of the photometric bands used here. An argument against this view is however provided by 
the ($J-H$)-($H-K_{S}$) diagram (Figure~\ref{figure:JH_HK}). When the suspected variable star (source 6)
and the high-temperature RCG sources (1, 28 and 14) are removed from the diagram, the distinction between 
RCG and HSB models becomes quite clear, with the HSB stars seeming to follow an extinction vector which is 
distinct from that followed by G/K giants. If the stars which are here classified as HSB were simply RCG stars with
different types of infrared excess, then such a distinct separation would not be expected. So it seems likely that 
the stars modelled as HSB type are distinct from G/K giants, although without spectroscopic identification of 
the stellar types, it cannot be confirmed that the hot-star plus bremsstrahlung model provides the correct
physical interpretation for these sources. 

It is also interest to examine whether the number of HSB sources detected here is consistent 
with the known frequency of early-type stars with bremsstrahlung emission. This work reveals six such 
sources in a survey area of 1.575~deg$^{2}$, i.e., at a surface 
density of about 4~deg$^{-2}$. Sources of this type are 
ubiquitous in studies of mid-IR excess: \citet{b20} report 126 such sources, which corresponds
to a surface density (in the GLIMPSE survey region) of about 0.6~deg$^{-2}$. In that work however, the 
typical $K_{S}$ magnitude of the early-type star involved is about 8. The work presented here is sensitive to 
$K_{S} \leq 12.9$. A crude extrapolation (based on a uniform spatial distribution of sources)
of the surface density of these sources from the work of \citet{b20} to 
the flux limits adopted here yields a value of order 100~sources~deg$^{-2}$. It does not seem unreasonable then that 
a small fraction of such sources fall within the region of the $K_{S}/(J-K_{S})$ colour magnitude diagram
used to isolate RCG stars.  

Turning now to the sources which are identified as being G/K giants with excesses at $8 ~\umu$m, a key 
question relates to the nature of the observed mid-IR excess. If the emission arises from a debris disc, 
it might be expected that the spectrum of the emission would be reasonably well approximated by a 
blackbody spectrum. Given that the photometric data presented here do not extend to wavelengths beyond
$8 ~\umu$m, it is not possible to determine a blackbody temperature of the component giving rise to 
the mid-IR excess. However, an estimate can be made on the upper limit to the black body temperature, 
by requiring that the excess component does not exceed measured values at $5.8 ~\umu$m. Inspection of
blackbody models in comparison to the SEDs of two strong excess sources (2 and 27, whose SEDs are shown in 
Figure~\ref{figure:SED_RCG}) reveals that models in which mid-IR emission arises from a
blackbody component are a good representation of the data when the blackbody temperatures are lower than 
about 450~K, and temperatures in excess of 600~K can clearly be ruled out. 

Another possible source of mid-IR emission is from PAHs. This is particularly pertinent to this study 
as one of the few G/K giants which has been subject to mid-IR spectroscopy, HD~233517 \citep{b13}, 
shows strong PAH emission. In the $8 ~\umu$m band, the infrared excess of this K2III star is 
dominated by PAH emission which has an amplitude which is approximately 70~percent of the photospheric emission
in this band (estimated from figure~1 of Jura et al.). The model used here to describe red-clump
giants with mid-IR excesses included a Gaussian component which was chosen to provide a reasonably good
representation of the $7.9 ~\umu$m PAH feature in the spectrum of HD~233517. I note that several sources 
have 8$ ~\umu$m excesses of about 0.5 magnitudes, which corresponds to the relative 
amplitude of Gaussian component with respect to the photospheric emission being similar to that 
observed in HD~233517. On the basis of these data I would not claim 
that PAH emission is the dominant cause of mid-IR excesses all G/K giants, but note that it is a possibility
which deserves further investigation. 

It is also of interest to examine the consequences of the measured incidence of mid-IR excesses in G/K giants in relation 
to the corresponding incidence in their main-sequence precursors. It is known that main-sequence stars give rise to 
mid-IR excesses which most likely arise from hot dust in debris discs.  Dust is produced by collisions within debris discs
but its emission depends on the radiation field of the host star. As a star with a disc evolves 
beyond the main-sequence, it is likely that the collisional processes are relatively unaffected, whereas the
radiation environment changes as the luminosity (and temperature) of the star change. An increase in luminosity 
of the host star between the main-sequence and the red-clump has two distinct effects. The first is that the 
blackbody temperature of dust (at a given radius from the star) will increase by a factor 
$(L_{\mbox{\scriptsize RC}}/L_{\mbox{\scriptsize MS}})^{1/4}$, where 
$L_{\mbox{\scriptsize MS}}$ and $L_{\mbox{\scriptsize RC}}$ are the stellar luminosities on the 
main-sequence and in the red-clump respectively. Secondly, if the lifetime ($t_{\mbox{\scriptsize pr}}$) of dust 
(i.e. from its generation to its accretion onto, or sublimation close to, the star)
is determined by Poynting-Robertson (P-R) drag, then this lifetime will vary according to 
\begin{equation}
\frac{t_{\mbox{\scriptsize pr,RC}}}{t_{\mbox{\scriptsize pr,MS}}}=\frac{L_{\mbox{\scriptsize MS}}}{L_{\mbox{\scriptsize RC}}}
\end{equation}
(where the subscripts MS and RC refer to main-sequence and red-clump respectively). 
While the assumption of the importance of P-R drag would be contrary to the theoretical expectation 
that dust lifetimes are determined by collisional timescales in massive debris discs \citep{b24}, it is useful to explore 
such a scenario in the light of the results presented in this paper. 

In order to analyse how these changes might have an impact on the observations presented here, 
it is necessary to identify 
the main-sequence precursors to red-clump stars. While recognising that 
red-clump stars, can in principle, have a relatively wide range of masses ($0.5 M_{\odot}$ to 
$2.5 M_{\odot}$, see e.g. \citealt{b26}) I choose a value of $2 M_{\odot}$ as being representative 
of the mass of the main-sequence precursor. This would correspond to spectral type A5V, with a luminosity of about 
$12 L_{\odot}$. If such a star evolves to form a canonical red-clump star, its luminosity would increase to about 
$46 L_{\odot}$.

The effect that this would have on dust temperature is that it would increase by about 40~per~cent. The
detection of a difference in mean temperatures between a population of main-sequence stars and red-clump stars
would require a much more comprehensive survey of mid-IR spectra than are currently available and certainly 
would require longer wavelength data than used in this study. Even if such a survey could be conducted,
the known variation in main-sequence disc temperatures, ranging from about 200~K to 800~K (Uzpen et al., 2007) would
make it very difficult to interpret such a study. 

On the other hand, the effect of variation of stellar luminosity on $t_{\mbox{\scriptsize pr}}$ may have more readily apparent
observable consequences. Under the assumption that dust lifetimes are dominated by P-R drag,
then provided that dust is generated by episodic events (such as in the solar system, e.g. \citealt{b08}),
and that the mean interval between dust generating events
is much greater than $t_{\mbox{\scriptsize pr}}$, then the incidence of mid-IR excesses will be proportional to $t_{\mbox{\scriptsize pr}}$. 
 
The observed incidence of mid-IR excesses in A type main-sequence stars is reported by \citet{b20}
as $1.0 \pm 0.5$ per~cent. Using the values given above for $L_{\mbox{\scriptsize MS}}$ and $L_{\mbox{\scriptsize RC}}$, 
the ratio of the decay timescales $t_{\mbox{\scriptsize pr,RG}}/t_{\mbox{\scriptsize pr,MS}}=0.27$, 
and within the scenario described above, this should correspond to 
the ratio of incidences of mid-IR excesses. Hence the expected incidence of mid-IR excesses in G/K giants would 
be expected to be $(2.7 \pm 1.4) \times 10^{-3}$. This is in agreement with our observed value of
$(1.8 \pm 0.4) \times 10^{-3}$. Hence, a model in which dust lifetimes in debris discs are dominated by P-R drag 
leads to a natural explanation of the observed incidence of mid-IR excesses in G/K giants. 

\section{Conclusions}
The major conclusion of this study is that mid-IR excesses are exhibited by G/K giants albeit at a very low
level of incidence. This could be consistent with the level of mid-IR excesses in their main-sequence 
precursors provided that dust generation is episodic and dust lifetimes are determined by P-R~drag rather than 
collisional processes. In passing, I note that this work has obtained a
value for the incidence of mid-IR excesses in G/K giants which is better determined than the corresponding
incidence in main-sequence stars. While it is assumed here that the mid-IR emission arises from relatively large dust grains in a 
debris disc, it should be emphasised that the nature of the mid-IR emission in this population remains uncertain, and requires 
mid-IR spectroscopic investigation. 

This study has illustrated that the technique of isolating samples of red-clump stars from CM diagrams
can be a useful technique to study this population of stars. I would expect that this technique could be put to good use by 
its application to a study of the incidence of excesses at longer infrared wavelengths.

\section*{Acknowledgments}
This publication makes use of data products from the Two Micron All Sky Survey, which is a joint project of the 
University of Massachusetts and the Infrared Processing and Analysis Center/California Institute of Technology, 
funded by the National Aeronautics and Space Administration and the National Science Foundation. I should like to 
thank Glenn White and the anonymous referee for comments which have helped to improve this paper.

\label{lastpage}

\clearpage
\pagestyle{empty}
\setcounter{table}{0}
\begin{landscape}
\begin{table}
\begin{center}{
\scriptsize
\begin{tabular}{rrrrrrrrrrrrrrrrrrrrrr}
\multicolumn{1}{c}{ID.}
   &    \multicolumn{1}{c}{Designation}   
                    &  \multicolumn{1}{c}{$J$}    
                            & \multicolumn{1}{c}{$\sigma (J)$}  
                                    & \multicolumn{1}{c}{$H$}  
                                           & \multicolumn{1}{c}{$\sigma (H)$} 
                                                   & \multicolumn{1}{c}{$K_{S}$}
                                                          & \multicolumn{1}{c}{$\sigma (K_{S})$}
                                                                 & \multicolumn{1}{c}{$[3.6]$} 
                                                                         &\multicolumn{1}{c}{$\sigma ([3.6])$}  
                                                                                & \multicolumn{1}{c}{$[4.5]$} 
                                                                                        & \multicolumn{1}{c}{$\sigma ([4.5])$}  
                                                                                               & \multicolumn{1}{c}{$[5.8]$}  
                                                                                                       & \multicolumn{1}{c}{$\sigma ([5.8])$}  
                                                                                                              & \multicolumn{1}{c}{$[8.0]$}  
                                                                                                                      & \multicolumn{1}{c}{$\sigma ([8.0])$} 
                                                                                                                              &\multicolumn{1}{c}{$E_{K8}$}
                                                                                                                                     & \multicolumn{1}{c}{SNR$_{EK8}$}
                                                                                                                                          \\ \hline
  1 & G045.1000+00.8808   & 13.759& 0.034& 12.739& 0.029& 12.214& 0.029& 11.745& 0.049& 11.688& 0.051& 11.451& 0.081& 11.484& 0.057&  0.280& 4.2\\
  2 & G045.1844+00.7612   & 14.202& 0.035& 13.173& 0.033& 12.764& 0.040& 12.390& 0.066& 12.272& 0.110& 12.036& 0.211& 11.371& 0.097&  0.983& 9.2\\
  3 & G045.2769+00.7123   & 14.212& 0.037& 13.105& 0.029& 12.763& 0.032& 12.397& 0.058& 12.438& 0.090& 12.223& 0.152& 11.797& 0.134&  0.552& 4.0\\
  4 & G045.2852+00.9988   & 14.242& 0.060& 13.179& 0.053& 12.798& 0.051& 12.523& 0.066& 12.430& 0.066& 12.518& 0.189& 11.881& 0.104&  0.505& 4.2\\
  5 & G045.4108+00.8768   & 13.589& 0.033& 12.608& 0.062& 12.328& 0.033& 12.062& 0.061& 12.135& 0.080& 11.886& 0.117& 11.577& 0.074&  0.406& 4.9\\
  6 & G045.6806+00.7488   & 14.323& 0.035& 13.365& 0.035& 12.872& 0.033& 11.644& 0.054& 11.487& 0.067& 11.371& 0.106& 10.997& 0.059&  1.460& 20.9\\
  7 & G046.2397+00.9526   & 12.690& 0.021& 11.837& 0.021& 11.529& 0.020& 11.245& 0.052& 11.231& 0.050& 11.233& 0.087& 10.896& 0.047&  0.325& 6.2\\
  8 & G046.3252+00.8710   & 14.369& 0.031& 13.257& 0.026& 12.871& 0.033& 12.442& 0.053& 12.399& 0.061& 12.439& 0.128& 12.011& 0.080&  0.428& 4.9\\
  9 & G046.6118+00.9688   & 13.640& 0.023& 12.580& 0.022& 12.175& 0.023& 11.842& 0.047& 11.883& 0.067& 11.598& 0.095& 11.484& 0.066&  0.271& 3.8\\
 10 & G046.6781+00.7324   & 14.045& 0.024& 13.152& 0.029& 12.474& 0.018& 11.314& 0.075& 10.736& 0.055& 10.231& 0.050&  9.645& 0.030&  2.370& 64.6\\
 11 & G046.6833+00.8797   & 14.187& 0.032& 13.069& 0.030& 12.577& 0.028& 12.274& 0.062& 12.291& 0.091& 12.057& 0.114& 11.776& 0.076&  0.327& 4.0\\
 12 & G046.7233+00.7512   & 14.333& 0.045& 13.217& 0.043& 12.750& 0.026& 12.524& 0.075& 12.458& 0.055& 12.222& 0.131& 11.814& 0.120&  0.472& 3.8\\
 13 & G046.9192+00.7417   & 12.700& 0.024& 11.983& 0.033& 11.542& 0.024& 10.921& 0.048& 10.728& 0.058& 10.544& 0.070& 10.337& 0.035&  0.899& 20.3\\
 14 & G047.5359+00.8104   & 12.701& 0.025& 11.867& 0.033& 11.595& 0.025& 11.210& 0.053& 11.097& 0.051& 11.221& 0.067& 11.037& 0.045&  0.271& 5.1\\
 15 & G047.8477+00.9287   & 13.454& 0.028& 12.408& 0.033& 12.031& 0.026& 11.864& 0.052& 11.952& 0.083& 11.648& 0.103& 11.148& 0.119&  0.479& 3.9\\
 16 & G048.1801+00.9573   & 14.089& 0.022& 12.964& 0.035& 12.625& 0.033& 12.253& 0.076& 12.362& 0.095& 12.140& 0.255& 11.804& 0.100&  0.401& 3.8\\
 17 & G048.3706+00.7076   & 13.977& 0.025& 13.133& 0.031& 12.569& 0.023& 11.985& 0.052& 11.702& 0.068& 11.277& 0.076& 11.202& 0.066&  0.968& 13.6\\
 18 & G048.3865+00.7629   & 14.454& 0.038& 13.438& 0.040& 12.792& 0.028& 12.330& 0.058& 11.861& 0.072& 11.509& 0.083& 11.155& 0.072&  1.144& 14.4\\
 19 & G048.3924+00.8103   & 14.562& 0.033& 13.433& 0.043& 12.876& 0.032& 12.430& 0.053& 12.466& 0.095& 12.093& 0.147& 11.273& 0.089&  1.101& 11.5\\
 20 & G048.4911+00.8260   & 14.201& 0.055& 13.114& 0.063& 12.641& 0.054& 12.200& 0.056& 12.423& 0.120& 12.147& 0.131& 11.592& 0.142&  0.594& 3.8\\
 21 & G048.5964+00.7855   & 14.337& 0.036& 13.192& 0.037& 12.752& 0.035& 12.368& 0.049& 12.411& 0.076& 12.205& 0.145& 11.871& 0.104&  0.417& 3.7\\
 22 & G048.5970+00.9849   & 14.024& 0.035& 13.068& 0.031& 12.708& 0.027& 12.492& 0.065& 12.448& 0.085& 12.199& 0.128& 11.938& 0.101&  0.405& 3.8\\
 23 & G048.6036+00.9284   & 14.306& 0.038& 13.320& 0.037& 12.638& 0.048& 11.984& 0.055& 11.746& 0.082& 11.456& 0.087& 11.189& 0.062&  0.954& 11.7\\
 24 & G049.0793+00.7610   & 13.365& 0.022& 12.438& 0.030& 12.166& 0.021& 12.000& 0.048& 11.979& 0.069& 11.825& 0.089& 11.559& 0.072&  0.285& 3.8\\
 25 & G049.1099+00.8232   & 13.995& 0.032& 13.000& 0.031& 12.606& 0.044& 12.327& 0.058& 12.408& 0.094& 12.128& 0.123& 11.775& 0.092&  0.439& 4.2\\
 26 & G049.4210+00.8476   & 13.558& 0.035& 12.816& 0.033& 12.316& 0.050& 11.808& 0.062& 11.450& 0.071& 11.204& 0.091& 11.032& 0.066&  0.946& 11.0\\
 27 & G049.5812+00.9666   & 14.158& 0.074& 13.094& 0.048& 12.708& 0.043& 12.406& 0.064& 12.501& 0.103& 12.191& 0.185& 11.120& 0.171&  1.174& 6.6\\
 28 & G049.9290+00.8456   & 13.804& 0.050& 12.772& 0.062& 12.241& 0.030& 11.729& 0.055& 11.512& 0.071& 11.546& 0.094& 11.343& 0.076&  0.442& 5.2\\
 29 & G050.1256+00.9808   & 13.990& 0.041& 13.083& 0.049& 12.458& 0.037& 11.712& 0.075& 11.417& 0.079& 11.148& 0.070& 10.679& 0.044&  1.334& 21.9
\end{tabular}
\normalsize
\caption{Photometric data for reliable sources selected as candidates for having $8 ~\umu$m excesses. Column 1 (ID.) is the 
identifier used to refer to individual sources in this paper. For each photometric band $\sigma()$ represent the single standard deviation uncertainties. 
The quantities $E_{K8}$ and SNR$_{EK8}$ are defined in the text.} 
\label{table:candidates}
}\end{center}
\end{table}

\end{landscape}
\clearpage


\end{document}